\documentclass[12pt]{article}
\usepackage{graphicx}
\newcommand{\sss}{\scriptscriptstyle}

\newcommand {\be}{\begin{equation}} 
\newcommand{\ee}{\end{equation}}    

\def\dds1{\frac{\partial}{\partial s_1}}

\def\vte{v_{{\sss T}e}}

\def\d{d\kern-0.8 ex\vrule height 1.3 ex depth-1.24 ex width 0.7 ex
\kern 0.15 ex}
\def\D{D\kern-1.7 ex\vrule height .87 ex depth-0.8 ex width 0.7 ex
\kern 0.95 ex}

\begin{document}
\baselineskip 20 pt

\begin{center}

\Large{\bf Slow EIT waves as  gravity modes }

\end{center}


\begin{center}

 J. Vranjes

{\em Belgian Institute for Space Aeronomy,  Ringlaan 3,
1180 Brussels, Belgium}

\end{center}

\vspace{2cm}

{\bf Abstract:}
 The  EIT waves  [named after the  Extreme-ultraviolet Imaging Telescope (EIT) onboard
the Solar and Heliospheric Observatory (SOHO)] are in the literature usually described as  fast magneto-acoustic (FMA) modes. However, observations show that a large percentage  of these events propagate with very slow speeds that may be as low as 20 km/s. This   is far below the FMA wave speed which  cannot  be below the sound speed, the latter being typically larger than  $10^2$ km/s in the corona. In the present study it is shown that, to account for such low propagation speed, a different wave model should be used, based on the theory of gravity waves, both internal (IG)  and surface (SG) ones. The  gravity modes are  physically completely different     from the FMA mode, as they are  essentially dispersive and in addition the IG wave is a transverse mode. Both the IG and the SG mode separately can provide proper propagation velocities in the whole low speed range.

\vspace{2cm}


\vspace{2cm}

\vfill
\pagebreak

{\bf I. INTRODUCTION }
\vspace{0.5cm}

One of the most striking details related to so called EIT waves is the  wide range of the  propagation velocity reported in the literature. Velocities of  50-60 km/s were reported in Ref.~1,  together with those of 350 km/s. In the same reference, values of 500 km/s are mentioned also as observed by Yohkoh/SXT. In Ref.~2 the values in the range 25-450 km/s are reported observed by SOHO EIT. Much more on that issue is available in Ref.~3, and also in Ref.~4 where velocities as low as  20 km/s are reported.

The EIT wave is usually  described in terms of the fast magneto-acoustic mode, which implies  the phase propagation speed given by
\be
v_p= \left(c_s^2 + c_a^2\right)^{1/2}, \quad c_s^2=\frac{\kappa (T_e+ T_i)}{m_i}, \quad c_a^2=\frac{B_0^2}{\mu_0 m_i n_0}.\label{vp}
\ee
 Clearly, the Alfv\'{e}n velocity $c_a$, determined by two plasma parameters gives some space to adjust to such a large span of  values, and even to provide the low values mentioned above.   For example, taking $n_0=10^{16}$ m$^{-1/3}$, $B_0=10^{-4}$ T, and
$n_0=3\cdot 10^{14}$ m$^{-1/3}$, $B_0=2\cdot 10^{-5}$ T, for $c_a$ yields $22$ km/s, and $25$ km/s, respectively. It remains questionable  though how realistic such parameters are. Even the `classic' value of the phase speed of the EIT waves, that is  around 250 km/s, implies  $n_0=2\cdot 10^{14}$ m$^{-1/3}$, $B_0=2\cdot 10^{-4}$ T, hence, a rather low value for the magnetic field. The impression is that, to account for such speeds, certain values of the plasma-$\beta$ are just  adopted   with the only purpose of fitting it into  the assumed and pre-selected fast magneto-acoustic wave model. However, according to Ref.~3, out of 160 studied events of the EIT waves,   more than 60 percent were  with the speed below $c_a$.

On the other hand, the FMA wave speed includes the sound speed too.  In calculating $c_s$  one has much less freedom for adjustment, because it is determined by the temperature only. For example, taking the ion and electron temperatures equal, and changing it in the range $0.6-2$ million K, one obtains $c_s$ in the range of $100-182$ km/s.  Thus, $c_s$ remains  above $10^2$ km/s, and no adjustment of the plasma-$\beta$ may help. According to Eq.~(\ref{vp})  there will  always  be $v_p\geq c_s$, therefore  the observed low values of $v_p$ mentioned above (just a few tens of km/s) can not possibly be explained within the fast magneto-acoustic wave model.

This  inspired some researchers to make a completely different models, claiming that the EIT perturbations are not waves at all.  Such non-wave models are merely focused on how the EIT  perturbations are produced in the first place, e.g., due to some specific change in the magnetic field topology. However, the origin of perturbations is not an issue here. From basic plasma theory we know that  an initial perturbation, no matter how produced, will propagate as a  wave, and  regardless of its  shape, it  can be decomposed into Fourier harmonics. Hence, one has to  deal with waves in any case.

      What remains as one option  is to search for some different wave model that could give proper propagation speed values in the low velocity range mentioned above.  This should by no means reduce  the validity of so widely accepted  fast magneto-acoustic (FMA) wave model, which indeed seems to  be able to capture many essential features of the fast EIT perturbations. Rather, it should represent a possible supplement in the slow speed domains where the FMA model appears to be not applicable.
Regarding the waves propagating orthogonal to the magnetic field, that seems to be the case with the EIT perturbations, the choice is very  limited, especially in view of the time and space scales involved in the perturbations.  Yet,  such plasma modes exist,  and those are the  internal gravity (IG) wave and the intermediate surface gravity (SG)  wave, which have features completely different as compared to the FMA mode. Those features of interest for the EIT waves will briefly be described in the following sections.

\vspace{0.5cm}
{\bf II. INTERNAL GRAVITY WAVES}
\vspace{0.5cm}

The physics of the IG wave is well known from standard books dealing with fluids  in external gravity field, as excellent examples to mention just the books  of Kundu$^5$ and Sturrock$^6$, where the mode is described in the frame of ordinary single fluid theory. In the present two-component plasma case applicable to  the coronal plasma, the  starting   equations for ions, in usual notation,  are the continuity, momentum and energy equation, respectively,
\be
\frac{\partial n_i}{\partial t} + \nabla\cdot(n_i \vec v_i)=0, \label{e1}
\ee
\be
m_i n_i \left(\frac{\partial }{\partial t} + \vec v_i\cdot \nabla\right)  \vec v_i=- e n_i \nabla \phi-\nabla p_i + m_i n_i \vec g,\label{e2}
\ee
\be
\left(\frac{\partial }{\partial t} + \vec v_i\cdot \nabla\right) \left(p_i n_i^{-\gamma}\right)=0. \label{e3}
\ee
Here, the gravity is assumed in the $z$-direction, $\vec g=-g\vec e_z$, the ions are singly-charged, and $\phi$ is the electrostatic potential.

 For low-frequency perturbations (with a phase speed much below the electron thermal velocity), electrons will be treated as isothermal, and their motion will be described by the quasi-neutrality condition $n_e=n_i$ (implying spatial scales far above the Debye length),  and by the momentum equation in the inertia-less limit
\be
0= e n_e \nabla \phi-\kappa T_e\nabla n_e + m_e n_e \vec g. \label{e4}
\ee
The magnetic field is neglected and a proper justification  for this will be given in the forthcoming text.

\noindent{\bf A. Equilibrium plasma}

Assuming a static and isothermal equilibrium without electric fields, and for a constant gravity, for the density  from Eqs.~(\ref{e2}, \ref{e4}) one obtains  the barometric formula
\be
n_j \sim \exp(-z/H),\label{e5a}
 \ee
 where $H=\kappa T_j/(m_j g)$ ($j=e, i$) is the characteristic scale length for the stratified medium. Note that the same expression can be obtained for an ordinary neutral gas.
 According to the formula (\ref{e5a}), there should be a difference (with height) in the density for species with different mass. In the case of electrons and ions this would then violate the quasi-neutrality. To preserve the quasi-neutrality,  the plasma will produce a macroscopic electric field $\vec E_0=E_0 \vec e_z$ [known also as the Pannekoek-Rosseland (PR) field$^{7,8}$ or ambipolar gravity-induced field] which opposes the gravity  in order to maintain its macroscopic quasi-neutrality. The electric field is oriented in such a way to lift the ions against the gravity,  and in the same time to pull  the electrons down. Hence, the appropriate momentum equations  for the two species will read
 \be
 e n_0  E_0 + \kappa T_0 \frac{dn_0}{dz} + m_e n_0 g=0, \quad E_0=-\nabla \phi_0,
 \label{e5}
 \ee
 \be
 e n_0 E_0 - \kappa T_0 \frac{dn_0}{dz} - m_i n_0 g=0.
 \label{e6}
 \ee
 Here, we used the quasi-neutrality $n_{i0}=n_{e0}=n_0$, and for simplicity  the same temperature for both species is assumed. Subtracting the two equations yields the PR field
 \be
 E_0=\frac{(m_i-m_e)g}{2 e}\simeq \frac{m_ig}{2 e}. \label{e7}
 \ee
 Setting this into Eqs.~(\ref{e5},\ref{e6}), contrary to Eq.~(\ref{e5a})  one now finds {\em the same} distribution for both electrons and ions
 \be
 n_0=N_0\exp\left(-\frac{z}{H_p}\right), \quad H_p=\frac{2 \kappa T_0}{m_i g}. \label{e8}
 \ee
  It is seen that  the plasma characteristic length $H_p$ is twice larger than $H$ for a neutral gas (obtained above also for ions without the effects of the PR field), and in the same time for electrons it is shorter than $H$ by a factor $2 m_e/m_i$.  These conclusions are general and
valid for any external gravity field and plasmas with two species.  Yet, these expressions are approximative because $g$ and $T_0$ are assumed locally constant with height $z$. An application of this effect to ion acoustic oscillations
 can be found in Refs.~9, 10. It can easily be shown that for two general species $a$ and $b$, with the charge $-q_a$ and $q_b$, and with different temperatures $T_a\neq T_b$, the PR field will read
 \be
 E_0=g \frac{m_b T_a - m_a T_b}{q_a T_b + q_b T_a}.\label{ef}
 \ee
Although frequently overlooked in the literature, this electric field is reality. The  field of the same origin  has been experimentally detected on the surface of metallic conductors.$^{11,12}$ It
appears due to the redistribution of electrons and nuclei in metallic objects placed in the gravity field.
 In the case of an expanding
atmosphere (e.g., in the solar wind)  the
magnitude of the  electric field can substantially be
greater than the Pannekoek-Rosseland value presented here. Details on that issue are available in Refs.~13, 14.

The presence of {\em several  charged ion species}  may drastically  change the previously obtained monotonous exponential distribution (\ref{e8}). To
demonstrate this, one may  assume the presence of $l$ singly charged ion species of equal temperature and write the
corresponding equilibrium momentum equations $-\kappa T \partial n_j/\partial z- n_jm_j g + en_j E=0$, $j=1, \cdots l$. Performing a summation of the
set of $l$ equations, together with the electron equation, with the help of quasi-neutrality,  $n_e=\sum_1^l
n_j$,  for the ambipolar electric field we obtain $E=\widehat{m}(z)g/2e$.  Here,
\be
\widehat{m}(z)=\sum_1^l m_j
n_j(z)/\sum_1^l n_j(z) \label{q9a}
\ee
 is an effective,  spatially dependent mean  ion mass. Using these expressions in the equation for  one  specific $j$-th
ion species, we obtain  $d(\ln n_j)/dz=-[m_j- \widehat{m}(z)/2] g/(\kappa T)$. It is seen that  in fact it {\em yields a
 growth with}  $z$  as long as $m_j< \widehat{m}(z)/2$. Hence, lighter ion species may have an inverse density distribution in lower layers. It may be concluded that  in general $H_j=H_j(z)$, and this dependence includes a possible change of the sign too. Such distributions may be expected in environments with dominant heavy ions, details related to  the terrestrial atmosphere  are available in Ref.~15. In the corona the situation is just the opposite, therefore the effects of extra ion species will not be studied here.
The presented picture of the PR field and the change in the scale length (in the case of equal temperatures enlarging it by a factor 2 for protons, and  in the same time reducing it by a factor  1000 for electrons) are based on the model of a quiet, stratified fluid. However, turbulence and mixing of different layers, especially in the lower solar atmosphere, may effectively reduce its importance.


\noindent {\bf B. Small perturbations}

Linearized momentum equations for the two species, with the help of Eqs.~(\ref{e5}-\ref{e8}), become
\be
m_i n_0 \frac{\partial \vec v_{i1}}{\partial t}= - e n_0 \nabla \phi_1 - \nabla p_{i1} - n_1 g m_i \frac{\tau}{1+ \tau} \vec e_z,
\label{i}
\ee
\be
0=  e n_0 \nabla \phi_1 - \kappa T_e \nabla n_1 - n_1 g m_i \frac{1}{1+ \tau} \vec e_z. \quad \tau=\frac{T_i}{T_e}.
\label{e}
\ee
Note that the gravity term in the electron equation contains the ion mass. The gravity terms in the two equations are equal if $\tau=1$, implying that the electrons and protons  effectively behave as particles of the same weight (but with different inertia, hence 0 on the left-hand side in the electron equation). This is the consequence of the PR field. The other two linearized equations are
\be
\frac{\partial n_1}{\partial t} + n_0 \nabla\cdot \vec v_{i1} + \vec v_{i1}\cdot \nabla n_0=0, \label{e10a}
\ee
\be
\frac{\partial p_1}{\partial t}  - \frac{\gamma p_0}{n_0} \frac{\partial n_1}{\partial t}  +  n_0^{\gamma} \left(\vec v_{i1}\cdot \nabla\right) \left(p_0 n_0^{-\gamma}\right)=0. \label{e10b}
\ee

To satisfy flux conservation, the perturbations propagating in the $(x, z)$-plane are taken of the form$^{6, 9, 10}$
\be
n_1=\widehat{n}_1\exp\left[-\frac{z}{2H} + i\left(k_x x+ k_z z-\omega t\right)\right],\label{a}
\ee
\be
p_1=\widehat{p}_1\exp\left[-\frac{z}{2H} + i\left(k_x x+ k_z z-\omega t\right)\right],\label{b}
\ee
\be
v_{(x,z)1}=\widehat{v}_{(x,z)1}\exp\left[\frac{z}{2H} + i\left(k_x x+ k_z z-\omega t\right)\right],\label{c}
\ee
%
%
\be
\phi_{1}=\widehat{\phi}_{1}\exp\left[\frac{z}{2H} + i\left(k_x x+ k_z z-\omega t\right)\right],\label{d2}
\ee
where $\widehat{\rho}_1$ etc., is used to denote the constant amplitude. Setting these expressions into
Eqs.~(\ref{i}-\ref{e10b})  yields the following equations for the amplitudes $\widehat{n}_1, \widehat{p}_1, \widehat{v}_{x1}, \widehat{v}_{z1}$:
\[
k_x \kappa T_e \widehat{n}_1 + k_x \widehat{p}_1 - \omega m_i N_0 \widehat{v}_{x1}=0,
\]
\[
\left[a m_i g + \left(i k_z + \frac{1}{2 H_p}\right) \kappa T_e\right] \widehat{n}_1 + \left(i k_z - \frac{1}{2 H_p}\right)\widehat{p}_1
\]
\[
 -
i \omega m_i N_0 \widehat{v}_{z1}=0,
\]
\[
-i \omega \widehat{n}_1 + i k_x N_0 \widehat{v}_{x1} + \left(i k_z - \frac{1}{2 H_p}\right) N_0 \widehat{v}_{z1}=0
\]
\[
- i \omega \gamma \kappa T_i \widehat{n}_1 - i \omega  \widehat{p}_1 + \frac{\kappa T_i N_0 (\gamma-1)}{H_p} \widehat{v}_{z1}=0.
\]
Here, $a=T_i/(T_e+ T_i)$, the number of equations was reduced by eliminating the electrostatic potential by using  the electron momentum equation. Its $x$-component  gave $n_1/n_0=e \phi_1/(\kappa T_e)$, while its $z$-component was identically equal to zero. Due to the shape of Eqs.~(\ref{a}-\ref{d2}), and after using Eq.~(\ref{e8}), the common exponential term cancels out. After a few steps this yields the
dispersion equation
\be
\omega^4-\left[\omega_c^2+ v_s^2 \left(k_x^2 + k_z^2\right)\right] \omega^2 + \omega_b^2 v_s^2 k_x^2=0. \label{e11}
\ee
Here,
\[
v_s^2=\frac{\kappa \left(T_e+ \gamma T_i\right)}{m_i}, \quad \omega_c=\frac{v_s m_i g}{2 \kappa (T_e+ T_i)}=\frac{v_s}{2H_p},
\]
\be
H_p=\frac{\kappa(T_e+ T_i)}{m_i g}, \quad \omega_b^2=\frac{(\gamma-1)m_i g^2 T_i}{\kappa (T_e+ T_i)(T_e+ \gamma T_i)}. \label{bv}
\ee
In $H_p$  the temperatures of the two species are kept different, $\omega_c$ is the ion acoustic (IA) wave  cut-off frequency, and $\omega_b$ is the buoyancy
[Brunt-V\"{a}is\"{a}l\"{a},  BV] frequency.

Eq.~(\ref{e11}) describes coupled the  IA and the internal gravity (IG) waves. The frequencies of the two modes are usually  well separated, for comparison see a detailed analysis for ordinary fluids in Ref.~6.  Hence, neglecting the last term in Eq.~(\ref{e11}) yields the IA mode
\be
\omega_{\sss{IA}}^2=\omega_c^2 + \left(k_x^2 + k_z^2\right) v_s^2. \label{e11a}
\ee
On the other hand, in the low frequency range one may neglect $\omega^4$ in Eq.~(\ref{e11}), which yields
\[
\omega^2= \frac{\omega_b^2 k_x^2 v_s^2}{\omega_c^2 + k^2 v_s^2}=\frac{\omega_b^2 k_x^2 v_s^2}{\omega_{\sss{IA}}^2}, \quad k^2=k_x^2 + k_z^2.
\]
In case when $\omega_c^2 \ll  k^2 v_s^2$ this yields the frequency for the IG mode presented  in Ref.~5 for  incompressible fluids and after using the Boussinesq approximation,
\be
\omega^2_{\sss{IG}}= \frac{\omega_b^2 k_x^2 }{ k_x^2 + k_z^2}. \label{e12}
\ee
According to conventional criterion of incompressibility, as long as the perturbed velocity $|v_1|$ is much below the sound velocity in the system, Eq.~(\ref{e12}) will be valid. In fact, obtaining   Eq.~(\ref{e12}) was the main reason for the preceding text.  Some  well known and interesting properties of the IG mode can easily be seen by analyzing Eq.~(\ref{e12}). The  phase velocity of the mode is
\be
\vec v_{\sss{IG}}=\frac{\omega}{k^2} \vec k=\frac{\omega_b k_x}{k^3}\left(k_x\vec e_x+ k_z \vec e_z\right).\label{e13}
\ee
On the other hand, its  group velocity is
\be
\vec c_{\sss{IG}}=\vec e_x\frac{\partial\omega}{\partial k_x}+ \vec e_z\frac{\partial\omega}{\partial k_z}=\frac{\omega_b k_z}{k^3}\left( k_z \vec e_x - k_x \vec e_z\right).\label{e14}
\ee
This yields
\be
\vec c_{\sss{IG}} \cdot \vec v_{\sss{IG}}=0. \label{e15}
\ee
From Eqs.~(\ref{e13}-\ref{e15}) it is seen that the horizontal components of $\vec c_{\sss{IG}}, \vec v_{\sss{IG}}$ are in the same direction, while their vertical components are equal by magnitude and opposite. In general, {\em the vectors of the two velocities  are perpendicular to each other}, implying that the mode is hydrodynamically transverse, i.e., the fluid elements oscillate parallel to the lines of constant phase. From Eq.~(\ref{e13}) it follows that the pure IG mode cannot propagate in the direction of the gravity vector (the phase velocity vanishes for $k_x=0$), while its group velocity  vanishes for $k_z=0$ (no energy transfer by the IG wave for its purely horizontal propagation). For an oblique propagation, if  the phase velocity has a positive $z$-component the corresponding group velocity component is negative, and vice versa. In the laboratory experiments with stratified liquids, e.g., a water with some vertical salinity profile  [see Ref.~5 and references cited therein], any vertical or horizontal displacement due to some source will cause oblique IG waves, with vertical group velocity component dependent on the frequency. It turns out that the more frequent oscillation of the source  the more vertical transfer of the wave energy. An interesting application of these features to the heating of chromosphere by the IG waves has been proposed long ago.$^{16}$ Very recent observations in fact confirm that  model,$^{17}$ showing that almost half of the required energy for the chromospheric heating comes from the IG waves launched  from the photosphere.

However, its  most important feature for the present study follows for a purely horizontal mode propagation.
 In such a case the perturbations propagate in the direction in which the medium is  isotropic, while their $z$-dependence  is self-consistently taken into account through Eqs. (\ref{a}-\ref{d2}), yet the two directions are not mixed,  the corresponding characteristic scale-lengths are practically independent and the common exponential term then cancels out.
In this limit from Eq.~(\ref{e12})  it is seen that $\omega=\omega_b$. Hence, the wave frequency achieves its maximum value, but more importantly, it becomes {\em independent on the wave number.} This further implies that, although the wave frequency is constant and strictly determined by the temperature and gravity  [see Eq.~(\ref{bv})], the wave phase speed is $v_{\sss{IG}}=\omega_b/k_x$. Being decoupled from the IA mode, the  horizontally propagating IG waves   are purely transverse, the actual displacement of the plasma due to the wave is in vertical direction,   and, regardless of their wave-lengths, their frequency is constant. Hence, physically the mode  is completely different from the fast magneto-acoustic wave which, in the presence of a vertical background magnetic field $\vec B_0=B_0 \vec e_z$,  implies the density compressions in {\em horizontal} direction and the compressive component of the magnetic field in  vertical $z$-direction.

Exactly because of such a geometry of the horizontally propagating  IG mode, and in particular because of the {\em vertical displacement} of the plasma (i.e., along the magnetic field vector),    the magnetic field was neglected from the beginning, because a plasma volume element moving up and down along the magnetic field lines due to the wave (propagating horizontally)  will not be affected by the field. Of course, such an idealized picture will hold only as long as  the frequencies of the two   modes  are well separated. Otherwise, there will be two physically rather different  modes (the IG and the fast magneto-acoustic one),   described by a   dispersion equation similar to  (\ref{e11}), but with the magnetic field effects included.


\vspace{0.5cm}
{\bf III. APPLICATION TO SLOW EIT WAVES}
\vspace{0.5cm}

To get some feeling about the quantities defined above, adopting $T_e=T_i=10^6$ K, $\gamma=5/3$, $g=275$ m/s$^2$, one finds $v_s=148$ km/s, $\omega_c=1.23\cdot 10^{-3}$ Hz, $\omega_b=1.07\cdot 10^{-3}$ Hz, and $H_p=6\cdot 10^7$ m. Note that the corresponding electron thermal velocity $\vte=4\cdot 10^6$ m/s justifies the earlier used inertia-less electron limit.

Because $\omega_b$ is so close to $\omega_c$, the two modes can only be  well separated  for large enough values of $k_x$. What `large enough' may mean is seen in Fig.~1, where the ratio of the internal gravity and ion acoustic wave  phase velocities $v_{\sss{IG}}/v_{\sss{IA}}=\omega_{\sss{IG}}/\omega_{\sss{IA}}=\omega_b/(\omega_c^2 + k_x^2 v_s^2)^{1/2}$ is given in terms of the wave number $k_x$ (multiplied by a factor $10^7$)  and for two  plasma  temperatures $T_e=T_i=T=1$ and $ 2$ million Kelvin. It is seen that the modes are well separated within a very large span of the wave-lengths that may be  of interest for the EIT modes. However for wavelengths of the order of $10^2$ Mm (far left side in Fig.~1) the coupling with the acoustic mode should be taken into account.

For the   value of  $k_x=4\cdot 10^{-8}$ m$^{-1}$ (i.e., $\lambda\simeq 150$ Mm),     from the graph in Fig.~1, the phase speed  $v_{\sss{IG}}$ becomes around 25 km/s. This perfectly describes the slow EIT perturbations. However, the constant frequency of the mode $\omega_b=1.07\cdot 10^{-3}$  Hz, implies perturbations with the wave period $T=2 \pi/\omega_b$ of 98 minutes, that look too long. The wave period can become lower for shorter $H_p$. In fact, the value for $H_p$ used above appears to be larger by a factor 2-3 from the values based on observations in coronal loops.$^{18}$. Hence, taking $H_p$ shorter by  a factor 3, the dispersion equation (\ref{e11}) becomes modified. After solving such a  new dispersion equation we find $\omega_b=0.0026$ Hz, and thus the wave period $T=2 \pi/\omega_b= 40$ minutes. In addition, accounting for the presence of 10 percent of helium and calculating the mean density, yields  the mean ion mass equal to 1.3 proton mass, and with this the  period can further be reduced to 35 minutes.

Clearly,  the buoyancy frequency and  the period of the IG wave are  sensitive to small variations of parameters. The value for the wave period presented here    is about the best that can be obtained from the present model. An additional improvement would be to include the magnetic field and to study the coupled fast magneto-acoustic (FMA) and IG modes. The actual displacements of the fluid due to these two modes (and this holds for the IA case above too) are in fact orthogonal to each other. So the simple picture of the transverse (vertical) incompressible plasma displacement due to the IG mode alone, will become much more complicated in the presence of the magnetic field and when the two modes are coupled. The coupling to the FMA mode will  make the IG  mode partly compressible too, and in the same time it will affect frequencies of both modes. This should particularly hold for a slightly  oblique propagation with respect to the vertical direction, when the pure IG mode  is described by Eq.~(\ref{e13}).    However, the magnetic field perturbations cannot be treated exactly within the present model. This is seen from the following. Having the magnetic field perturbed implies the use of the Amp\`{e}re law
\be
 \nabla\times \vec B_1=\mu_0 \vec j_1= \mu_0 e n_0\left(\vec v_{i1}- \vec v_{e1}\right). \label{am}
 \ee
 In view of the earlier introduced variation of the velocity (\ref{c}), and for the given equilibrium density variation (\ref{e8}), this would  require that the variation of $B_1$ is the same as the variation of the density (\ref{a}), $\sim \exp(-z/2H)$. However,  one is supposed to use the Faraday law too, $\nabla\times \vec E=-\partial \vec B/\partial t$,  where the variation of $B_1$ in the $z$-direction should be the same as the variation of the electric field potential (\ref{d2}), i.e.,  $\sim \exp(z/2H)$. Both are not possible, hence there can be no common  exponential term to  cancel out, and the given analytical model does not work. This case  can only  be studied numerically.


\vspace{0.5cm}
{\bf IV. SURFACE GRAVITY WAVE}
\vspace{0.5cm}

 The same procedure for obtaining the slow propagation velocity of the EIT waves within the IG wave scheme,  can  be successfully extended to completely fill in the gap between the values obtained above (around  25 km/s in the given example), and the minimum possible velocity  (i.e., $c_s$) that  would follow from the fast magneto-acoustic wave model [which is $\geq 10^2$ km/s, see Eq.~(\ref{vp})]. Though this works only for the reduced $H_p$ (by a factor 3), and for  the mean ion mass equal to 1.3 proton mass,  as explained in Sec.~3. The maximum wave-lengths   are below 200 Mm and  wave periods are around 35 minutes.

  However, even without such modifications, i.e., keeping the original $H_p$ and $m_i$,  the gap can be filled in also  by using  the surface gravity (SG) waves  in fluids with intermediate depths (comparable to the wave-lengths),  by identifying the scale length $H_p$ with the fluid depth. Performing derivations similar to Sec.~2,  the frequency of the gravity wave becomes$^5$
  \[
    \omega_{\sss{SG}}=[g k_x \tanh( k_x H_p)]^{1/2}.
     \]
 It should be stressed that a  similar model is successfully used also in the description of terrestrial Rossby and gravity waves, where the bulk  Earth's atmosphere is approximated   by a layer of the thickness  $h$ obtained in the same manner as the quantities $H$ and $H_p$ earlier in the text.

The phase velocity $v_{\sss{SG}}=\omega_{\sss{SG}}/k_x $ of the surface  gravity mode  is given in Fig.~2 in the range of   wave-lengths of interest for the EIT waves, for $m_i$ equal to the proton mass,  for  the temperature of the plasma $T_e=T_i=10^6$ K, and $H_p=6 \cdot 10^7$ m as earlier. Here, the wave-length is taken in the interval $0.1 - 160$ Mm, yielding  the wave periods of the SG wave in the interval between 0.8   and 32 minutes, respectively. Hence, the surface gravity mode alone is capable of yielding proper propagation speeds for the EIT waves in the whole low velocity range.
  Regarding the coupling with the IA  waves, similar arguments as in the previous text hold here too.  However, because SG wave includes both vertical and horizontal plasma displacement, the coupling with the FMA wave is expected to be much more pronounced and it requires numerical simulations.

\vfill
\pagebreak

\vspace{0.5cm}
{\bf V. CONCLUSIONS}
\vspace{0.5cm}

The present study is focused on the slow range of the propagation speed of the EIT perturbations, i.e., those that are well below the magneto-acoustic velocity.   Such low velocities, that do not fit into the fast magneto-acoustic wave description of the EIT perturbations, in the recent past  have been a major argument against the wave nature of those events.

In this study, basic plasma and fluid theory is used in order to show that such slow propagating perturbations can successfully be described within the theory of gravity waves. In fact two possible alternatives are suggested, based on the internal and surface gravity wave theory.  Hence, these results should at least partly reestablish and support the wave-based model of the EIT perturbations.

In its simplest form, the internal gravity mode  described in the text does not produce density variation. As already mentioned before, their coupling to the FMA mode in fact may include such variation. On the other hand, the horizontally propagating IG  mode implies vertical displacements (in the $z$-direction) of the plasma, which means  that layers of different density (with $z$) will be mixed together. For example,  a denser plasma volume raised  (within one half-length of the wave propagating in the $x$-direction)  from some level  $z_0$ to $z_1= z_0+ \delta z_0$,  will be surrounded by a less dense plasma corresponding to the level $z_1$, and vice versa in the other half-length. An  observer will see a propagating perturbation  accompanied by density variations, although the mode itself is not compressible in the first place.

The surface gravity  mode  is rather different, as it includes both horizontal and vertical displacements, and its group and phase velocities are in the same direction.    Analysis of the fluid motion inside the wave$^{5,19,20}$ shows  that in the wave crest the fluid motion is generally in the direction of the wave, while in the troughs it is opposite. For a horizontal propagation the group velocity is non-zero and the mode transfers energy, which is not the case with  the IG mode. This  all should perhaps be a possible recipe for observations, to eventually distinguish the SG mode  from both the IG and FMA  modes.  Though as a reaction to sudden perturbations (like flares), the coronal  plasma can, at any time, support different  plasma waves that on the other hand may become more or less coupled. Lack of  ideal conditions, like for example the absence of a clear surface, implies that in a realistic situation  some hybrid gravity mode may be expected that would  include features of  both  gravity modes discussed in the text. The velocity, and time and space scales of the resulting gravity mode should be between the two. A proper  description of   such waves and especially  their coupling to FMA mode requires  numerical simulations.


\section*{Acknowledgments}
This work has partly received funding  from    the Solar-Terrestrial Center of Excellence/SIDC, Royal Observatory of Belgium,  Brussels.

\vfill
\pagebreak

{\bf Figure captions:}

\begin{description}
\item{Fig. 1:} Ratio of the phase velocities (and frequencies) of the internal gravity mode and the ion acoustic mode, in terms of the wave-number.

\item{Fig. 2:} The phase velocity $v_{\sss{SG}}=\omega_{\sss{SG}}/k_x $ of the surface  gravity mode in terms of the wave-length.
 
 \end{description}

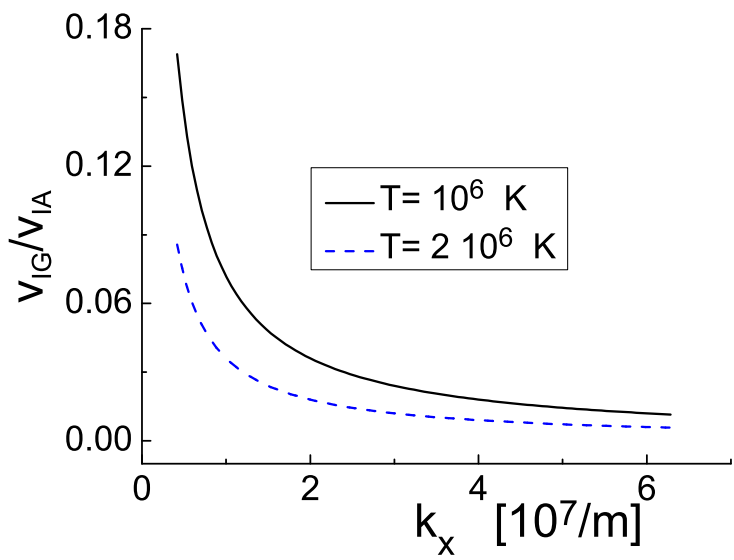
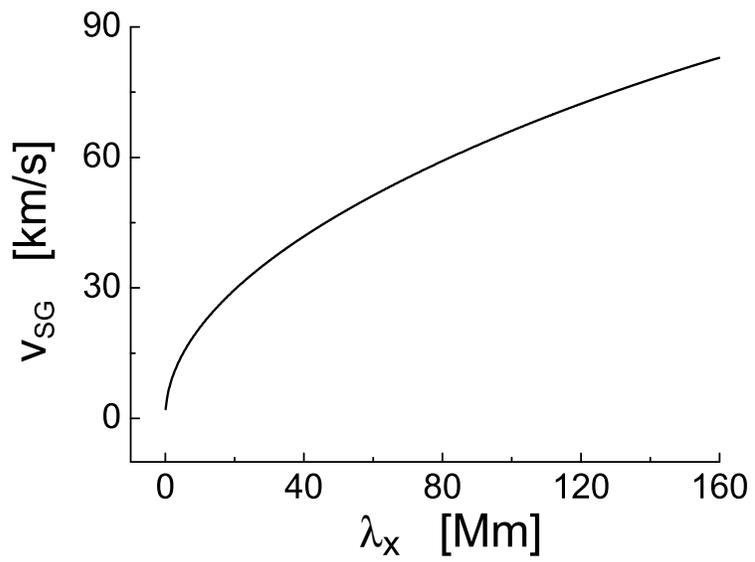
\end{document}